\documentclass[11pt,preprintnumbers,amsmath,amssymb,floatfix,showpacs]{revtex4}
\usepackage{graphicx,epsfig,latexsym,amssymb}
\usepackage{multirow,amsmath,array,booktabs}
\usepackage{longtable}
\usepackage[section]{placeins}
\usepackage{bm}
\usepackage{color}
\usepackage{epstopdf}
\usepackage{ulem}
\usepackage{mathrsfs}
\usepackage{subfigure}
\usepackage{slashbox}
\begin{document}

\title{Isospin-density dependent pairing from infinite nuclear matter to finite nuclei}
\email{zss76@buaa.edu.cn, lisheng.geng@buaa.edu.cn, caolg@bnu.edu.cn}

\author{Xu Meng}
\affiliation{\mbox{$^{1}$ School of Physics, Beihang University, Beijing 100191, China}\\}
\author{Shisheng Zhang}
\affiliation{\mbox{$^{1}$ School of Physics, Beihang University, Beijing 100191, China}\\}
\author{Lin Guo}
\affiliation{\mbox{$^{1}$ School of Physics, Beihang University, Beijing 100191, China}\\}
\author{Lisheng Geng}
\affiliation{mbox{$^{1}$ School of Physics, Beihang University, Beijing 100191, China}\\}
\affiliation{\mbox{$^{2}$ Beijing Key Laboratory of Advanced Nuclear Materials and Physics,
Beihang University, Beijing 100191, China}\\}
\affiliation{\mbox{$^{3}$ Beijing Advanced Innovation Center for Big Data-Based Precision Medicine,
School of Medicine and Engineering, Beihang University, Beijing, 100191}\\}
\affiliation{\mbox{$^{4}$School of Physics and Microelectronics, Zhengzhou University, Zhengzhou, Henan 450001, China}\\}
\author{Ligang Cao}
\affiliation{\mbox{$^{5}$ College of Science and Technology, Beijing Normal University, Beijing 100875, China}\\}
\begin{abstract}
The effective isospin-density dependent pairing interaction (P1)
[S. S. Zhang, U. Lombardo and E. G. Zhao, Sci. Chin. Phys. Mech. Astro. {\bf 54}, 236 (2011)]
extracted from neutron pairing gaps for $^1$S$_0$
in asymmetric nuclear matter calculations 
[S. S. Zhang, L. G. Cao, U. Lombardo, et al. Phys. Rev. C {\bf 81}, 044313 (2010)]
is employed to study the bulk properties of Ca, Ni, Zr and Sn isotopes.
The odd-even mass (OEM) staggering is calculated by the Skyrme Hartree-Fock plus BCS method (SHF + BCS) with the SkP interaction.
For comparison, we study two other types of isovector effective pairing interactions.
One is also extracted from pairing gaps of infinite nuclear matter by the
Brueckner-Hartree-Fock (BHF) method but for free spectrum (P2).
The other is obtained by fitting the empirical OEM (P3).
An isoscalar effective pairing interaction (P4) is also adopted which is determined by fitting the empirical OEM.
We find that interaction P1 can better describe the OEM staggering of Ni, Zr and Sn isotopes
by $14.3 \%, 41 \%, 30.4 \%$ compared with interaction P2,
in terms of root mean square deviations to the empirical OEM, respectively.
On the other hand, the performance of P1 and P2 is comparable for Ca isotopes.
For Ca and Ni isotopes, P1 behaves similarly as P3,
but for Zr isotopes P1 is better than P3  by $\sim 34 \%$.
Among the four pairing interactions studied, P4 performs the worst.
Therefore, one may conclude that
for neutron pairings in finite nuclei, the isovector pairings are preferred than the isoscalar one.
It is quite interesting to note that the pairing interaction P1 extracted from nuclear matter calculations
can describe pairing gaps of finite nuclei as well as or even better than the interaction P3 directly fitted to finite nuclei.\\
\noindent{{\bf Keywords:} nuclear matter, pairing correlation, odd-even mass staggering, finite nuclei}
\end{abstract}

\pacs{21.10.Dr, 21.30.Fe, 21.60.-n}


\maketitle


\section{Introduction}\label{sec:intro}
\label{sec:introduction}
Pairing correlations play an important role in describing many observables and processes in nuclear physics, for instance,
the odd-even mass (OEM) staggering for finite nuclei~\cite{Zhang2011_SCPMA54-236,Zhang2010_PRC81-044313}, the superfluidity and cooling of neutron stars~\cite{Dean2003_RMP75-607,Dong2016_TAJ817-6}, r-process nucleosynthesis~\cite{Zhang2015_PRC91-045802}, etc.

Tremendous efforts have been made to extract pairing interactions from different observables.
One way is to fit the OEM staggering of finite nuclei via the empirical three-point or five-point pairing gap formulae with the experimental binding energies as inputs~\cite{Bertulani2009_PRC80-027303,Goriely2009_PRL102-152503}.
But pairing gaps can be different for three-point and five-point formulae,
and sometimes cannot reproduce small pairing for nuclei with (double) magic numbers or truly reflect the experimental OEM difference
since energy density functionals for odd-A systems are not as good as those for even-even systems~\cite{Hinohara2016_PRL116-152502}.
Recently, there is a proposal that nucleonic pairing can be extracted from nuclear density functional theory for pairing rotational bands in even-even nuclei with the Quasi-particle Random Phase Approximation (QRPA) method~\cite{Hinohara2016_PRL116-152502}.
A separable force of finite range is widely used to describe pairing correlations in normal nuclei~\cite{Tianyuan2009_PLB676-44}
and has recently been applied to provide effective pairing interactions for hyperons~\cite{Rong2020_PLB807-135533}.
Another alternative is to extract pairing interactions from asymmetric nuclear matter (ANM) calculations with the microscopic Brueckner-Hartree-Fock (BHF) method~\cite{Zhang2011_SCPMA54-236, Margueron2008_PRC77-054309, Chamel2008_NPA812-72, Zhang2010_PRC81-044313, Zhang2016_PRC93-044329},
and adopt local density approximation to obtain the isospin-density dependent parameters for finite nuclei.
In the last decade, Margueron, Sagawa and Hagino introduced zero-range isospin-density dependent effective pairing interactions~\cite{Margueron2007_PRC76-064316, Margueron2008_PRC77-054309}
by fitting to the corresponding pairing gaps of symmetric nuclear matter (SNM) and pure neutron matter (PNM)
obtained by the BHF method with and without medium polarization effect~\cite{Cao2006_PRC74-064301}.
But those pairing gaps presented in Ref.~\cite{Margueron2007_PRC76-064316} are extracted from free spectrum instead of the mean field
spectrum as clarified in our previous paper~\cite{Zhang2011_SCPMA54-236}.
In that paper, we proposed a new effective pairing interaction~\cite{Zhang2011_SCPMA54-236} from the mean field spectrum based on self consistent calculations, denoted by P1 in the following.

In this paper, we aim at applying the new pairing interaction P1 to describe finite nuclei
and to see whether one can provide a universal description of pairing correlations in nuclear matter
and finite nuclei. For comparison, we also study three other pairing interactions.
And we label the pairing interaction of Ref.~\cite{Margueron2008_PRC77-054309}
as P2, which is obtained in the same way as P1 but fitted to pairing gaps of free spectrum instead of the mean field.
We also consider a second isovector pairing interaction, referred to as P3, which is extracted from fitting to
the experimental OEM using the empirical three-point formula~\cite{Bertulani2009_PRC80-027303}.
In addition to the above two isovector types of pairing interactions,
we also study an isoscalar pairing interaction, denoted as P4, to check the impact of isospin dependence.

For the mean field part, we use the \verb+ev8+ code~\cite{Bonche2005_CPC171-49,Ryssens2015_CPC187-175}
with the SkP force. The particle-particle channel is described by the BCS approximation with four different pairing interactions detailed above.
Experimental binding energies are taken from AME2016~\cite{Wang2017_ChinPhysC41-030003}.

The paper is organized as follows. In Sec. II, we give a brief introduction of
the Skyrme Hartree-Fock  plus BCS method (SHF + BCS) and
describe the isospin-density dependent  and isoscalar pairing interactions.
Numerical details are also presented in this section.
Then, we take Ca, Ni, Zr and Sn isotopes as examples to compare the results of four effective pairing interactions in Sec. III.
Discussions on the corresponding potentials are analyzed as well.
Finally, we make a brief summary in Sec. IV.

\section{Theoretical Framework} \vspace*{-1mm}
\label{sec:theoretical}
In this section, we briefly review the SHF + BCS method, describe
 the isospin-density dependent pairing interactions, and spell out some numerical details.
\subsection{Skyrme force}
The Skyrme force is widely used in  Hartree-Fock calculations.
Its energy density functional~\cite{sly4} contains eight terms
\begin{equation} \label{eq:Skyrme}
        \mathcal{H}=\mathcal{K}+\mathcal{H}_0+\mathcal{H}_3+\mathcal{H}_{\rm eff}+\mathcal{H}_{\rm fin}+\mathcal{H}_{\rm so}+\mathcal{H}_{\rm sg}+\mathcal{H}_{\rm Coul}.\\
\end{equation}
They are the kinetic energy term $\mathcal{K}=(\hbar^2/2m)\tau$, zero range term $\mathcal{H}_0$, density dependent term $\mathcal{H}_3$,  effective mass term $\mathcal{H}_{\rm eff}$, finite range term $\mathcal{H}_{\rm fin}$, spin-orbit term $\mathcal{H}_{\rm so}$,
and $\mathcal{H}_{\rm sg}$ due to the tensor coupling with spin and its gradient.

These terms have the explicit expressions as follows,
\begin{equation}
\begin{split}
                \mathcal{H}_0=&\frac{1}{4}t_0\left[\left(2+x_0\right)\rho^2-\left(2x_0+1\right)\left(\rho_{\rm p}^2+\rho_{\rm n}^2\right)\right],\\
                \mathcal{H}_3=&\frac{1}{24}t_3\rho^\alpha\left[\left(2+x_3\right)\rho^2-\left(2x_3+1\right)\left(\rho_{\rm p}^2+\rho_{\rm n}^2\right)\right],\\
        \mathcal{H}_{\rm eff}=&\frac{1}{8}\left[t_1\left(2+x_1\right)+t_2\left(2+x_2\right)\right]\tau\rho\\
                              &+\frac{1}{8}\left[t_2\left(2x_2+1\right)-t_1\left(2x_1+1\right)\right]\left(\tau_{\rm p}\rho_{\rm p}+\tau_{\rm n}\rho_{\rm n}\right),\\
        \mathcal{H}_{\rm fin}=&\frac{1}{32}\left[3t_1\left(2+x_1\right)-t_2\left(2+x_2\right)\right]\left(\nabla\rho\right)^2\\
                              &-\frac{1}{32}\left[3t_1\left(2x_1+1\right)+t_2\left(2x_2+1\right)\right]\left[\left(\nabla\rho_{\rm p}\right)^2+\left(\nabla\rho_{\rm n}\right)^2\right],\\
         \mathcal{H}_{\rm so}=&\frac{1}{2}W_0\left[\vec{J}\cdot\nabla\rho+\vec{J}_{\rm p}\cdot\nabla\rho_{\rm p}+\vec{J}_{\rm n}\cdot\nabla\rho_{\rm n}\right],\\
         \mathcal{H}_{\rm sg}=&-\frac{1}{16}\left(t_1x_1+t_2x_2\right)\vec{J}^2+\frac{1}{16}\left(t_1-t_2\right)\left[\vec{J}_{\rm p}^2+\vec{J}_{\rm n}^2\right],
\end{split}
\end{equation}
in which $t_0$, $t_1$, $t_2$, $t_3$, $x_0$, $x_1$, $x_2$, $x_3$, and $W_0$ are fitted parameters, and $\rho=\rho_{\rm p}+\rho_{\rm n}$, $\tau=\tau_{\rm p}+\tau_{\rm n}$ are matter and kinetic densities, $\vec{J} = \vec{J}_{\rm p}+\vec{J}_{\rm n}$ is the spin-orbit current \cite{sly4}. We adopt the SkP force~\cite{Dobaczewski1984_NPA422-103} in our present study,
which is obtained by paying particular attention to pairing properties and accurate description of binding energies.

\subsection{Isospin-density dependent pairing interactions}

As a linear interpolation of the particle-particle interaction between symmetric nuclear matter and pure neutron matter,
 isospin-density dependent zero-range effective interactions are derived in Refs.~\cite{Margueron2007_PRC76-064316,Margueron2008_PRC77-054309}, for neutrons and protons, respectively,
\begin{equation} \label{eq:isospinpair}
\begin{split}
g_{\rm n}(\rho,\beta)&=1-\eta_s(\rho/\rho_0)^{\alpha_s}(1-\beta)-\eta_n(\rho/\rho_0)^{\alpha_n}\beta,\\
g_{\rm p}(\rho,\beta)&=1-\eta_s(\rho/\rho_0)^{\alpha_s}(1+\beta)+\eta_n(\rho/\rho_0)^{\alpha_n}\beta,\\
\end{split}
\end{equation}
where the four parameters $\eta_s, \eta_n, \alpha_s$ and $\alpha_n$ are adjusted to reproduce the exact
values of the pairing gaps in infinite nuclear matter.
The saturation density $\rho_0$ of the SNM is $\rho_0=0.16\ {\rm fm}^{-3}$ and
asymmetric parameter $\beta$ is defined as $\beta=(N-Z)/A$ or $\beta = (\rho_{\rm n}-\rho_{\rm p})/(\rho_{\rm n}+\rho_{\rm p})$,
in which $N$ ($Z$) is neutron (proton) number, $A = N + Z$ is mass number and $\rho_{\rm n}$ ($\rho_{\rm p}$) refers to neutron (proton) density.

The isoscalar pairing interaction reads,
\begin{equation} \label{eq:densitypair}
        g(\rho,\beta=0)=1-\eta_s(\rho/\rho_0)^{\alpha_s},\\
\end{equation}
where $\eta_s=1$ and $\alpha_s=1$.

In the BCS approximation~\cite{schul}, the pairing matrix element reads,
\begin{equation}
        \bar{v}^{\rm pair}_{q,k\bar{k}m\bar{m}}=-V_q\int{\rm d}^3r\ g_q(\rho,\beta)\Psi^\dagger_k(\bm{r})\Psi^\dagger_{\bar{k}}(\bm{r})\Psi_m(\bm{r})\Psi_{\bar{m}}(\bm{r}),\\
\end{equation}
where $q$ stands for n or p, $V_q = V_0$ is the pairing strength determined by the scattering length,
which reproduces the phase shift in the low energy region for a given cutoff energy,
$g(\rho,\beta)$ from Eq.~(\ref{eq:isospinpair}) or Eq.~(\ref{eq:densitypair}) is the form factor of the isovector or isoscalar pairing interaction, and $\Psi_k(\bm{r})$ is the wave function of the $k$th HF single-particle (s.p.) level \cite{Krieger1990_NPA517-275}.

\subsection{Pairing energy, binding energy and pairing gap in the SHF + BCS method}

In the SHF+BCS method, the binding energy $B$ of a nucleus can be written as a sum of five parts \cite{Ryssens2015_CPC187-175},
\begin{equation}
        B=-E_{\rm kin}-E_{\rm Sk}-E_{\rm Coul}-E_{\rm pair}-E_{\rm corr},\\
\end{equation}
where $E_{\rm kin}$ is the kinetic energy, $E_{\rm Sk}$ is the Skyrme energy, $E_{\rm Coul}$ is the Coulomb energy, $E_{\rm pair}$ is the pairing energy, and $E_{\rm corr}$ is the center-of-mass correction energy. $E_{\rm pair}$ is the main contribution of the pairing interaction to the binding energy. Other parts of the binding energy, such as $E_{\rm Sk}$, are functions of the density $\rho$, and the density $\rho$ is a function of the occupation probability $v_k^2$ of single-particle levels. Therefore, the binding energy will also be influenced by the pairing interaction through $v_k^2$.

In particular, the pairing energy $E_{\rm pair}$ can be written in the canonical basis as
\begin{equation}
        E_{\rm pair}=\sum_{k,m>0}f_k u_k v_k f_m u_m v_m \bar{v}^{\rm pair}_{k\bar{k}m\bar{m}},\\
\end{equation}
in which $u_k^2+v_k^2=1$, $\bar{v}^{\rm pair}_{k\bar{k}m\bar{m}}$ is the pairing matrix element, and
\begin{equation} \label{eq:cutoff}
        f_k=[1+e^{(\epsilon_k-\lambda_q-E_C)/\mu_q}]^{-1/2}[1+e^{(\lambda_q-\epsilon_k-E_C)/\mu_q}]^{-1/2},\\
\end{equation}
is the cutoff factor, where $\epsilon_k$ is the energy of the $k$th s.p. level, $\lambda$ is the Fermi energy, $E_C$ is the truncation energy of the pairing interaction, and $\mu_q$ is fixed to be 0.5 MeV \cite{Krieger1990_NPA517-275}.

The set of equations that determine the occupation probability of single-particle states $v_k^2$ are derived from the variation of
\begin{equation}
        \frac{\delta}{\delta v_j}\Big(2\sum_{k>0}\epsilon_kv_k^2-\lambda_q\langle\hat{N}_q\rangle+E_{\rm pair}\Big)=0,\\
\end{equation}
where $\lambda_q$ is the Lagrange multiplier, which is introduced to obtain the requested mean number of protons and neutrons.

The probability of the s.p. state $|k\rangle$ and its time-reversal s.p. state $|\bar{k}\rangle$ being occupied by one pair of neutrons or protons can be expressed as
\begin{equation}
        v_k^2=\big[1-(\epsilon_k-\lambda_q)/\sqrt{(\epsilon_k-\lambda_q)^2+f_k^2\Delta_{k\bar{k}}^2}\ \big]/2,\\
\end{equation}
in which
\begin{equation}
        \Delta_{k\bar{k}}=\sum_{m>0}f_m u_m v_m \bar{v}^{\rm pair}_{k\bar{k}m\bar{m}}\\
\end{equation}
is the pairing gap of the $k$th single-particle level.

For odd-$A$ nuclei, we use the blocking method of Ref. \cite{Bender2000_EPJA8-59} to consider the odd particle. When a pair of s.p states $|k\rangle$ or $|\bar{k}\rangle$ is chosen to be occupied or blocked, the occupied or unoccupied probability in the BCS state is fixed to be $u_k^2=v_k^2=1/2$, and the pairing gap of that s.p. levels is fixed to be $\Delta_k=0$ MeV.

\subsection{\label{sec:num}Numerical details}\vspace*{-1mm}

The ev8 code solves the HF+BCS equations to obtain binding energies iteratively with the imaginary time step method~\cite{Davies1980_NPA342-111}.
For even-even and odd-$A$ nuclei, we use the empirical three-point formula to extract the OEM staggering of isotopes,
\begin{equation} \label{eq:3point}
        \Delta_n^{(3)}(N,Z)=-\frac{\pi_N}{2}[B(N+1,Z)-2B(N,Z)+B(N-1,Z)],\\
\end{equation}
where $B(N,Z)$ is the binding energy of a nucleus, and $\pi_N$ is the parity of the isotope with neutron number $N$.

We adopt three isospin-density dependent pairing interactions: P1~\cite{Zhang2011_SCPMA54-236}, P2~\cite{Margueron2007_PRC76-064316,Margueron2008_PRC77-054309}, and P3~\cite{Bertulani2009_PRC80-027303}.
For comparison, the isoscalar interaction P4~\cite{Bertulani2009_PRC80-027303} is also considered.
We list the parameters of the pairing interactions in Table ~\ref{tab:parameter}.
One can see that P1 and P2  have the same cut-off energy $E_{C} = 40 $ MeV,
while P3 and P4  have a smaller cut-off energy  $E_{C} = 5 $ MeV.
As we know, effective pairing interactions are sensitive to the values of the energy (or momentum) cut-off (see, e.g., Fig.4 of Ref.~\cite{Zhang2010_PRC81-044313} for PNM).
Therefore, the parameters are quite different for P2 and P3, especially the potential strength $V_0$.
This will be further analyzed in the following section.

Before large-scale calculations, we check the convergence of OEM staggering with respect to the number of neutron wave functions.
This is necessary because the calculated OEM staggering of finite nuclei might not be right
if the basis space of wave functions is not large enough.
In the ev8 code, the number of neutron wave functions,``nwaven'', is an input parameter.
We choose neutron-rich Ca, Ni, Zr and Sn isotopes with magic or sub-magic proton number,
to calculate OEM staggering using the three-point formulas with the P1 pairing interaction.
We can see from Table ~\ref{tab:nwaven} that neutron pairing gaps $\Delta_n$ converge to a certain value
with the number of neutron wave functions increasing.
The absolute value of the differences for $\Delta_n$ from $N = 90$ to $N = 100$ represented by $\delta$ is
smaller than 0.03 MeV for all the cases, which guarantees the convergence.
In our later study, $N=100$ is used.

In most calculations, an absolute accuracy of $10^{-7}$ MeV can be achieved for binding energies.
Since the binding energies are at the order of 1000 MeV, the relative accuracy is better than  $10^{-10}$.
However, for some odd-$A$ nuclei, it is difficult to achieve such an accuracy.
Therefore, the criterion of convergence is fixed to be $10^{-5}$ MeV, which corresponds to an relative accuracy $10^{-8}$.
\begin{table}[htbp]
\caption{Parameters of the pairing interactions studied in the present work, in which P1, P2, and P3 are of isovector type,
and P4 is of isoscalar type. $E_C$ is the truncated energy, $V_0$ is the pairing strength,
$\eta_s$, $\alpha_s$, $\eta_n$, and $\alpha_n$ are the parameters of Eq.~(\ref{eq:isospinpair}).}
\label{tab:parameter}
\renewcommand{\arraystretch}{1}
\begin{ruledtabular}
\begin{tabular}{cccccc}
    Parameters&P1&P2&P3&P4\\
    \hline
    $E_C$ (MeV)&40&40&5&5\\

    $V_0$ (MeV fm$^3$)&542&542&824&1400\\

    $\eta_s$&0.729&0.664&0.677&1\\

    $\alpha_s$&0.522&0.522&0.365&1\\

    $\eta_n$&1.010&1.010&0.931&-\\

    $\alpha_n$&0.525&0.525&0.378&-\\
\end{tabular}
\end{ruledtabular}
\end{table}

\begin{table}[htbp]
\caption{Neutron pairing gaps $\Delta_{\rm n}$ for Ca, Ni, Zr and Sn isotopes as
a function of the number of neutron wave functions for the pairing interaction P1.
The absolute values of the difference for $\Delta_n$ from $N = 90$ to $N = 100$ are represented by $\delta$.
All the quantities are in units of MeV.}
\label{tab:nwaven}
\renewcommand{\arraystretch}{1}
\begin{ruledtabular}
\begin{tabular}{cccccc}
    &$N=80$&$N=90$&$N=100$&$\delta$\\
    \hline
    $\Delta_{\rm n}(^{52}{\rm Ca})$&1.03&0.95&0.95&0.00\\
    $\Delta_{\rm n}(^{74}{\rm Ni})$&1.23&1.24&1.22&0.02\\
    $\Delta_{\rm n}(^{110}{\rm Zr})$&1.40&1.37&1.40&0.03\\
    $\Delta_{\rm n}(^{116}{\rm Sn})$&1.36&1.38&1.38&0.00\\
\end{tabular}
\end{ruledtabular}
\end{table}

\begin{figure}[htbp]
    \includegraphics[width=1.15\textwidth]{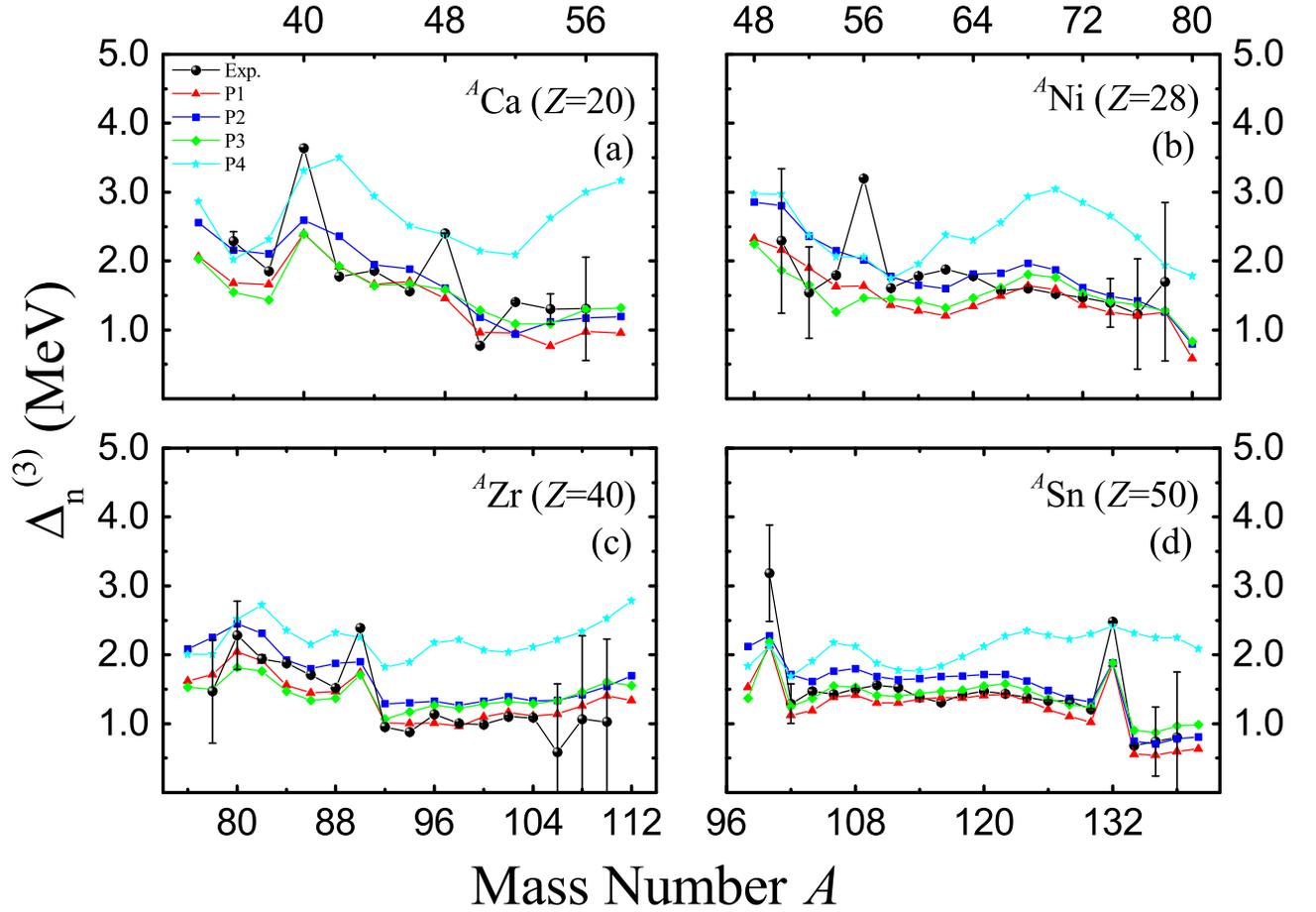}
    \caption{Odd-even mass staggering as a function of mass number A for Ca, Ni, Zr and Sn isotopes with the SkP force.
    P1 (solid red lines with triangles), P2 (solid dark blue lines with squares),
    and P3 (solid green lines with diamonds) are isovector pairing interactions
    and P4 (solid light blue lines with stars) is isoscalar one.
    The black solid circles with error bars
    correspond to experimental data.}
    \label{fig:OEM}
\end{figure}

\section{Results and discussions}

We calculate the binding energies of even-even and odd-$A$ Ca, Ni, Zr and Sn isotopes with the SkP  force,
then derive the OEM staggering via the three-point formula of Eq.~(\ref{eq:3point}).
It is known that pairing gaps obtained from the OEM staggering for those isotopes with magic or semi-magic number
$N = 20$, 28, 40, 50 and 82 are not reliable,
such as $^{40}$Ca, $^{48}$Ca, $^{56}$Ni, $^{78}$Ni, $^{90}$Zr, $^{100}$Sn and $^{132}$Sn.
Therefor, we ignore the neutron pairing gaps $\Delta_n$ for these nuclei in the later analysis.

In Fig.~\ref{fig:OEM}, we show the OEM staggering as a function of mass number A for Ca, Ni, Zr and Sn isotopes.
The solid red (dark blue) lines with triangles (squares) refer to the neutron pairing gaps obtained with the isovector pairing interactions P1 (P2),
which is extracted from the BHF + BCS calculations for SNM and PNM with the mean field (free) spectrum.
The solid green lines with diamonds correspond to the predictions of the isovector pairing interaction P3,
which is obtained by fitting to the experimental OEM staggering.
While the solid light blue lines with stars denote the results obtained with the isoscalar pairing interaction P4.
The black solid circles with error bars labelled by Exp. represent the experimental data.
Generally speaking, the neutron pairing gaps $\Delta_{\rm n}^{(3)}$
obtained with the isovector pairing interactions P1, P2 and P3,
are much closer to the experimental data than those from the isoscalar pairing interaction P4.
Therefore, the isospin-dependence of the pairing interaction seems to be crucial to reproduce the experimental OEM staggering.

To quantitatively evaluate the deviations of the theoretical predictions from the experimental OEM staggering,
we calculate the Root Mean Square Error (RMSE) of the neutron pairing gaps for all the isovector pairing interactions,
except for magic or semi-magic nuclei as mentioned above.
The RMSE is defined by
 \begin{equation} \label{eq:RMSE}
        \sigma=\sqrt{\frac{\sum\limits_{i=1}^N (\Delta_{i,{\rm cal}}-\Delta_{i,{\rm exp}})^2}{N}},\quad i=1,2,\cdots,N\\
\end{equation}
where $\Delta_{i,{\rm cal}}$ and $\Delta_{i,{\rm exp}}$ are the calculated and
the experimental OEM staggering of an isotope labeled by $i$, and $N$ is the number of isotopes considered.
We list the RMSEs in Table ~\ref{tab:RMSE} for a better understanding of
the predictions of different pairing interactions in comparison with the experimental data.
As mentioned above, the neutron pairing gaps for magic nuclei
$^{40}$Ca, $^{48}$Ca, $^{56}$Ni, $^{78}$Ni, $^{90}$Zr, $^{100}$Sn and $^{132}$Sn are omitted in calculating the RMSEs.

For Ni and Zr isotopes, P1 turns out to be the best among the isovector pairing interactions.
The RMSEs of the OEM staggering for Ni, Zr and Sn isotopes are reduced by $14.3 \%$, $41.0 \%$, and $30.4 \%$ respectively,
compared with those of P2,
which is also extracted from infinite nuclear matter with the BHF method but for free spectrum, instead of the mean field spectrum.
Moreover, the predictions of P1 for Ca isotopes are almost the same as those of P2,
and are better than those of P3 by $\sim 5 \%$. One should note that
 P3 is obtained by fitting to the experimental OEM staggering.
In addition,  for Zr isotopes P1 can reduce the RMSEs of OEM staggering
by $\sim 34.3 \%$ compared with P3.
For Sn isotopes, P1 can also give a better description of OEM staggering than P2 by $30 \%$, but not as good as P3.
It seems to be consistent with the conclusion of Ref.~\cite{Bertulani2009_PRC80-027303} that
 P3 is particularly good for Sn isotopes together with the SkP force. As for P4,
 it is much worse than the other three isovector pairings as we noted earlier.  Again,
 we stress that that neutron pairing gaps from infinite nuclear matter can be a good constraint
for constructing neutron pairing interactions for finite nuclei.

\begin{figure}[htbp]
    \centering
    \includegraphics[width=0.8\textwidth]{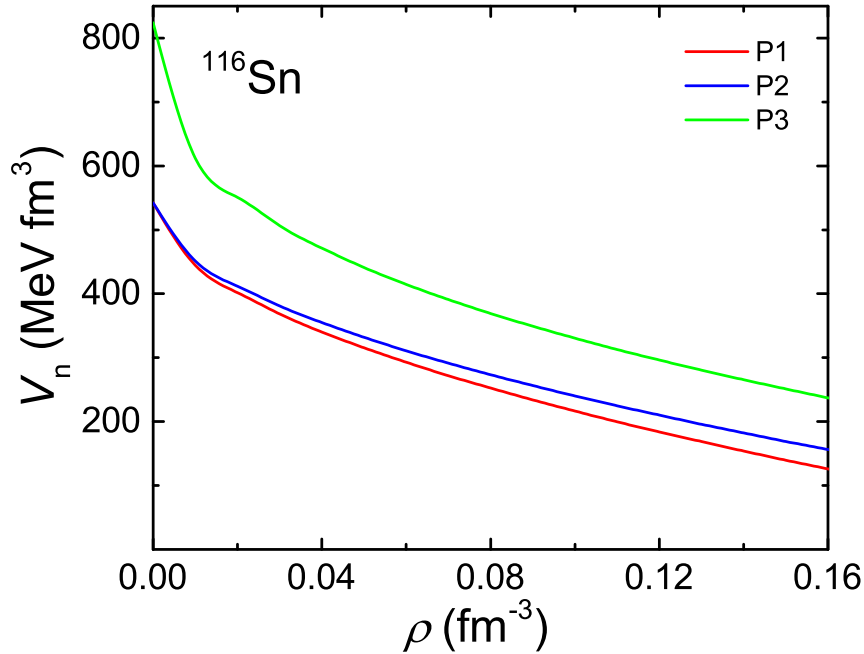}
    \caption{Neutron pairing interaction $V_{\rm n}=V_0\ g_{\rm n}(\rho,\beta)$ as a function of density $\rho$. $g_{\rm n}(\rho,\beta)$ is the neutron pairing form factor calculated from Eq.~(\ref{eq:isospinpair}). Parameters of the pairing interactions are listed in Table ~\ref{tab:parameter}.}
    \label{fig:V_rho}
\end{figure}

To further investigate the difference among the three isovector pairing interactions,
we simply choose $^{116}$Sn as an example to plot in Fig.~\ref{fig:V_rho} the neutron pairing interaction $V_{\rm n}=V_0\ g_{\rm n}(\rho,\beta)$
as a function of density $\rho$ with fixed asymmetric parameter $\beta=(66-50)/116\approx0.138$.
For increasing $\rho$, $V_{\rm n}$ decreases monotonically.
The curve of P3 is much different from those of P1 and P2,
due to the different energy cutoff.
On the other hand, P1 and P2 look similar.
Suppose that finite nuclei have the saturation density $\rho = 0.16 $ fm $^{-3}$, then
 $V_{\rm n}$ of P2 is larger than that of P1.
Correspondingly, the pairing gaps obtained with P2 should also be larger than those obtained with P1 in Fig.~\ref{fig:OEM}.
The main difference between P1 and P2 is the value of $\eta_s$, see Table ~\ref{tab:parameter},
$\eta_s$=0.664 for P2 is slightly smaller than $\eta_s$=0.729 for P1.
Since $\eta_s$ is  negative in  $g_{\rm n}(\rho,\beta)$ [see Eq.~(\ref{eq:isospinpair})],
smaller $\eta_s$ of P2 results in a larger $V_{\rm n}$ compared with P1.
Quite remarkably, it seems that pairing gaps in ANM impose strong constraint on those in finite nuclei,
to the extent that the same pairing interactions can be used for both cases.
More studies are needed to further corroborate such a conclusion.

To show the predictive power of the pairing interactions,
we extend our calculations to neutron rich nuclei $^{58}$Ca, $^{80}$Ni, $^{112}$Zr, $^{140}$Sn,
which are plotted in Fig.~\ref{fig:OEM} as well.
We have also calculated the OEM scattering of Ca and Sn isotopes with the SLy4 force~\cite{sly4}, and found that it
does not predict the pairing gaps as well as the SkP force.
Generally speaking, it underestimates the pairing gaps compared with measured data
and doubles the RMSEs than the SkP force for the three isovector pairing interactions.
Therefore, it justifies the use of SkP force in the SHF+BCS model to check the validity of pairing interactions
and it confirms the claim of Ref.~\cite{Dobaczewski1984_NPA422-103} that SkP can give better descriptions of pairing gaps.

\begin{table}[htbp]
\caption{RMSE of the OEM staggering of Ca, Ni, Zr and Sn isotopes in units of MeV and the relative errors.
The OEM staggering for magic nuclei $^{40}$Ca, $^{48}$Ca, $^{56}$Ni, $^{78}$Ni, $^{90}$Zr, $^{100}$Sn and $^{132}$Sn
are neglected in the calculation of RMSE.}
\label{tab:RMSE}
\renewcommand{\arraystretch}{1}
\begin{ruledtabular}
\begin{tabular}{lcccccccc}
    &P1&P2&$\dfrac{P2-P1}{P2}$&P3&$\dfrac{P3-P1}{P3}$&P4&$\dfrac{P4-P1}{P4}$ \\

    \hline
    Ca      &0.35&0.33&-6.1\%&0.37&5.4\%&1.17&70.1\%\\
    Ni      &0.30&0.35&14.3\%&0.30&0.0\% &0.94&68.1\%\\
    Zr      &0.23&0.39&41.0\%&0.35&34.3\%&1.00&77.0\%\\
    Sn      &0.16&0.23&30.4\%&0.12&-33.3\%&0.89&82.0\%\\

\end{tabular}
\end{ruledtabular}
\end{table}

\section{Summary} \vspace*{-1mm}
\label{sec:summary}

We studied the OEM staggering of Ca, Ni, Zr and Sn isotopes with the effective pairing interaction P1,
together with other three types of pairing interactions for comparison,
using the Skyrme Hartree-Fock  plus BCS method with the SkP force.
We showed that P1 is suitable for the description of OEM staggering in these isotopes,
especially, much better than the other pairing interactions for Ni and Zr isotopes.
For example, the pairing gaps for Zr isotopes are $41 \%$ ($34 \%$) better than those obtained with P2 (P3), accordingly.
Our predictions for Ca isotopes are comparable with those of P2 and P3,
which is reasonable since the parameters of P3 are obtained by fitting to the experimental OEM staggering~\cite{Bertulani2009_PRC80-027303}.
For Sn isotopes, the predictions of P1 are almost $\sim 30 \%$ better than those of P2,
and are comparable with the results of P3.
It is quite interesting to find that although the isospin-density dependent pairing interaction P1 is extracted from
the bare interaction for nuclear matter within the framework of the BHF method,
it can give reasonable description of the OEM staggering for Ca, Ni and Zr isotopes ($30 \%$ better than the results obtained with P2).
The isoscalar pairing P4 is also considered and turns out to be not very good,
which shows the importance of the isospin effect in pairing interactions.

One should note that the pairing interaction P1 is obtained by fitting to the pairing gaps
of SNM and PNM with the bare interaction in the framework of the microscopic BHF method.
It is interesting to see that without any tunable parameters,
it can give pretty good description of the OEM staggering compared with measured data.
For some isotopes, e.g. Ca and Zr isotopes (or Ni isotopes),
the  predictions are even better (or comparable) than those of P3,
extracted by fitting  to the experimental pairing gaps.
From this point of view, it is reasonable to say that neutron pairing gaps of infinite nuclear matter
can be a good constraint of the neutron pairing interaction in finite nuclei. In the future,
our ansatz could also be tested for proton pairing gaps.

\section{Acknowledgements}
{Discussions with Prof. Hiroshima Watanabe and Prof. Hagino are gratefully acknowledged.
Authors show their great thanks to Prof. Shan-Gui Zhou for his careful reading of this article.
This work was supported partially by the
National Natural Science Foundation of China under Grant
No.~11775014, No.~11975096, No.~11735003, No.~11975041, and No.~11961141004.}

\end{document}